\documentclass{PoS}

\newcommand{\BDlnu}{$B\to D \ell \nu$}
\newcommand{\BDstlnu}{$B\to D^* \ell \nu$}

\title{Semileptonic decays $B\rightarrow D^{(*)}l\nu$  at nonzero recoil}

\ShortTitle{Semileptonic decays $B\rightarrow D^{(*)}l\nu$ at nonzero recoil}

\author{\speaker{Si-Wei Qiu}%
        \thanks{Present address: Laboratory of Biological Modeling, NIDDK, NIH, Bethesda, MD 20892-5621, USA}\\
        Department of Physics and Astronomy, University of Utah, Salt Lake City, UT 84112-0830, USA\\
        E-mail: \email{siwei.qiu@nih.edu}}
\author{Carleton DeTar\\
        Department of Physics and Astronomy, University of Utah, Salt Lake City, UT 84112-0830, USA\\
        E-mail: \email{detar@physics.utah.edu}}
\author{Aida X.~El-Khadra\\
        Physics Department, University of Illinois, Urbana, IL 61801-3080, USA and \\
        Fermi National Accelerator Laboratory, Batavia, IL 60510-5011, USA}
\author{Andreas S.\ Kronfeld\\
        Fermi National Accelerator Laboratory, Batavia, IL 60510-5011, USA}
\author{Jack Laiho%
\thanks{Present address: Department of Physics, Syracuse University, Syracuse, NY 13244-1130, USA}\\
        SUPA, Department of Physics and Astronomy, University of Glasgow, Glasgow, G12~8QQ, United Kingdom}
\author{Ruth S. Van de Water\\
        Fermi National Accelerator Laboratory, Batavia, IL 60510-5011, USA}
\author{(Fermilab Lattice and MILC Collaborations)}

\abstract{We have analyzed the semileptonic decays $B\rightarrow
  D\ell\nu$ and $B\rightarrow D^*\ell\nu$ on the full suite of MILC
  (2+1)-flavor asqtad ensembles with lattice spacings as small as
  0.045 fm and light-to-strange-quark mass ratios as low as 1/20. We
  use the Fermilab interpretation of the clover action for heavy
  valence quarks and the asqtad action for light valence quarks. We
  compute the hadronic form factors for $B\rightarrow D$ at both zero
  and nonzero recoil and for $B\rightarrow D^*$ at zero recoil. We
  report our results for $|V_{cb}|$.}

\FullConference{31st International Symposium on Lattice Field Theory - LATTICE 2013\\
		July 29 - August 3, 2013\\
		Mainz, Germany}

\begin{document}
\section{ Introduction}
\label{sec:In}

The CKM matrix element $|V_{cb}|$ plays a prominent role in tests of
the Standard Model.  It normalizes the legs of the unitarity triangle.
The dominant uncertainty in $|V_{cb}|$ has come from theoretical
determinations of the decay rates for $B \to c\ell\nu + \ldots{}$.
The purpose of this work is to study the exclusive processes
\BDlnu\ and \BDstlnu\ in lattice gauge theory in order to reduce the
uncertainty in $|V_{cb}|$ determined from these decays.

The Fermilab Lattice and MILC collaborations are completing a $B$- and
$D$-physics program based on fourteen large ensembles of gauge
configurations, generated in the presence of (2+1)-flavors of improved
staggered (asqtad) sea quarks.  The strange-quark mass $m_s$ is kept
at approximately its physical value and the degenerate light (up and
down) quark mass $\hat m^\prime$ takes on values from $\hat
m^\prime/m_s = 0.05$ to 0.4.  The lattice spacings in these ensembles
range from approximately 0.045 fm to 0.15 fm, as discussed in
\cite{Qiu:2011ur}.  Clover (Fermilab) fermions are employed
for the bottom and charm quarks and staggered (asqtad) fermions, for the
light valence quarks with masses set equal to the sea quarks.
Among the quantities calculated are the hadronic form factors for
\BDlnu\ at nonzero recoil and for \BDstlnu\ at zero recoil

Our methodology for \BDlnu\ has been outlined in previous conferences
in this series \cite{Qiu:2011ur}, and for \BDstlnu, we
update our previously published results
\cite{Bernard:2008dn} with data from the
same full set of asqtad ensembles. For both decays, papers with details
are in preparation \cite{B2Dpaper,B2Dstarpaper}.  Here we focus on
details of the \BDlnu\ analysis that have not been reported
previously.

To avoid biases, the analysis described here was blind, following now
common practice in experimental high energy physics.  The vector
current renormalization constants were determined by two members of
the collaboration and reported to the rest of the collaboration,
multiplied by a common, secret blinding factor.  Only after the value
of $|V_{cb}|$ was obtained was the blinding factor revealed and
removed.  That is the value reported here.

\section{Notation}
\label{sec:Note}

We start by reviewing our notation.  The differential decay rate for
the exclusive process \BDlnu\ is given by
\begin{eqnarray}
    \frac{d\Gamma}{dw}(B\rightarrow D\ell\nu)=|\bar \eta_{\rm EW}|^2\frac
    {G_F^2|V_{cb}^2| M_B^5}{48\pi^3}(w^2-1)^{3/2}r^3(1+r)^2\mathcal{G}(w)^2\,\,.
\label{eq:diffdecay}
\end{eqnarray}
in the approximation that the masses of the leptons $\ell = e, \mu,
\nu_e, \nu_\mu$ are much smaller than the $B$ and $D$ mass difference
$M_B - M_D$.  The recoil parameter, $w = v\cdot v^\prime$, is the dot
product of the four-velocities of the $B$ and $D$ mesons, $v =
p_B/M_B$ and $v^\prime = p_D/M_D$, respectively, and $r = M_D/M_B$.
The factor $|\bar \eta_{\rm EW}|^2$ accounts for electroweak corrections.
The hadronic form factor $\mathcal{G}(w)$ is proportional to the
vector form factor $f_+(w) = (1+r)\mathcal{G}(w)/(4r)$.
%
%\begin{equation}
%    f_+(w)^2=\frac{(1+r)^2}{4r}\mathcal{G}(w)^2\,\,.
%\end{equation}
%
That form factor is obtained from a tensor decomposition of the
hadronic vector-current matrix element for the transition,
\begin{equation}
    \langle D(p_D) | \mathcal{V}^\mu | B(p_B) \rangle =
        f_+(q^2) \left[ (p_B+p_D)^\mu - \frac{M_B^2-M_D^2}{q^2}q^\mu \right]  +
        f_0(q^2) \frac{M_B^2-M_D^2}{q^2} q^\mu \, ,
\end{equation}
where the four-momentum transfer is $q=p_B-p_D$.  Here
$\mathcal{V}^\mu = \bar b \gamma^\mu c$ is the $b \to c$ vector
current and $f_+$ and $f_0$ are the vector and scalar form factors,
respectively.  

The alternative form factors $h_+$ and $h_-$ are convenient for lattice simulations:
\begin{equation}
    \frac{\langle D(p_D) | \mathcal{V}^\mu | B(p_B) \rangle }{ \sqrt{M_B M_D}} =
        h_+(w)(v + v^\prime)^\mu  + h_-(w) (v-v^\prime)^\mu \, ,
\end{equation}
They are related to $f_+$ and $f_0$ through
\begin{equation}
    f_+(q^2) = \frac{1}{2\sqrt{r}} \left[ (1+r) h_+(w) - (1-r) h_-(w) \right]; \,\,
    f_0(q^2) = \sqrt{r} \left [ \frac{w+1}{1+r} h_+(w) - \frac{w-1}{1-r} h_-(w) \right]
\label{eq:f_from_h}
\end{equation}

Lattice calculations at zero recoil ($w = 1$) typically have the
smallest errors. However, because of the phase space suppression near
zero recoil in $ B \rightarrow D \ell \nu$, evident from the factor
$(w^2-1)^{3/2}$ in Eq.~(\ref{eq:diffdecay}), experimental errors are
largest there.  Thus, we aim to work at nonzero recoil where the
combined experimental and theoretical error is minimized.

\section{Determination of the form factors}

Using procedures outlined in previous reports
\cite{Qiu:2011ur}, the hadronic form factors $h_+(w)$ and
$h_-(w)$ were determined from fits to appropriate three-point and
two-point correlation functions.  Some $w$-dependent adjustment was necessary
because the simulation values of the charm and bottom quark masses
($\kappa_b$ and $\kappa_c$) were slightly different from our final,
preferred, tuned values of these masses.  Statistical errors with
correlations were propagated through the entire calculation using a
single-elimination jackknife.

\begin{figure}
\centerline{
\includegraphics[width=0.5\textwidth]{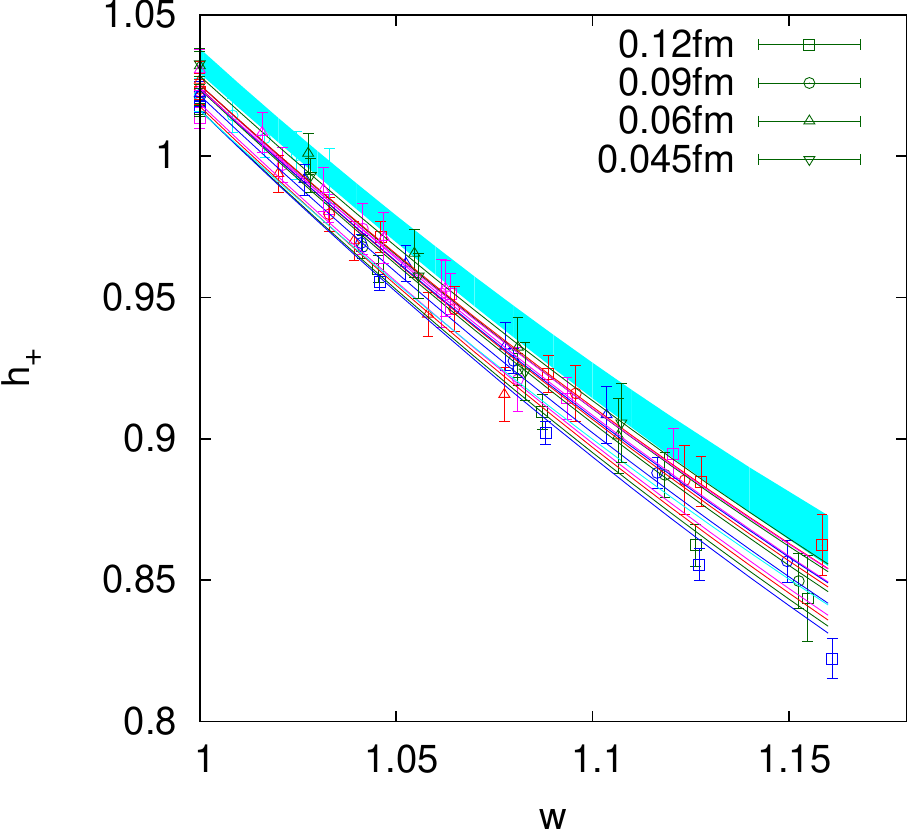} \hfill
\includegraphics[width=0.5\textwidth]{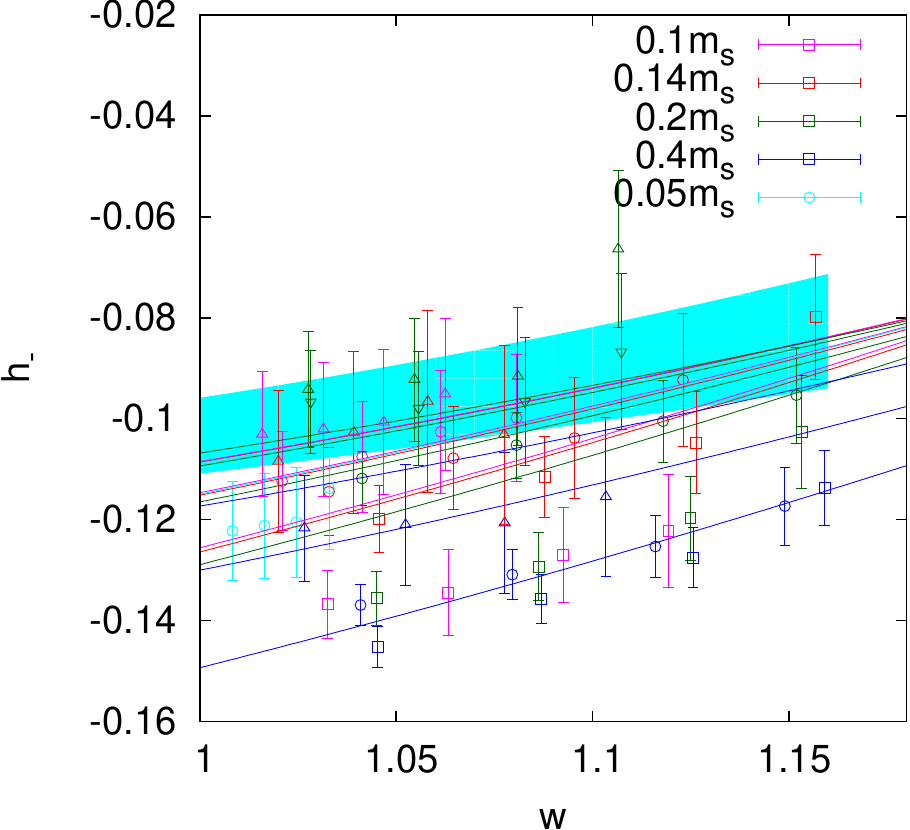}
}
\caption{
Global fit of all data for the form factors $h_+$ (left) and $h_-$ (right) vs $w$, the recoil parameter.
    The simultaneous fit gives $p = 0.53$.
    The blue band shows the physical continuum prediction.  In both plots the symbol
    shapes distinguish the lattice spacing as indicated in the left legend and the
    colors distinguish sea quark mass ratios, as indicated in the right legend.
    Errors are statistical only.
}
\label{fig:h+h-}
\end{figure}

Results for all ensembles are plotted in Fig.~\ref{fig:h+h-}. To
extrapolate to the physical quark mass and zero lattice spacing, we
use the following fit Ans\"atze for the chiral/continuum extrapolation
\begin{eqnarray}
    h_+(a,\hat m^\prime,w)     & =     & 1 + \frac{ X_+(\Lambda_\chi)}{m_c^2}
     - \rho_+^2 (w-1) + k_+ (w-1)^2
     + c_{1,+} x_l + c_{a,+} x_{a^2} + c_{a,w,+} x_{a^2}(w-1) + \nonumber \\
   & + & 
    c_{a,a,+} x_{a^2}^2 + c_{a,m,+} x_l x_{a^2} + c_{2,+} x_l^2 + 
    \frac{g_{D^*D\pi}^2}{16 \pi^2 f^2} {\rm logs}_{\rm 1-loop}(\Lambda_\chi,w,\hat m^\prime,a)
    \label{eq:chiral}\\
    h_-(a,\hat m^\prime,w)     &=      & \frac{X_-}{m_c} - \rho_-^2 (w-1) + k_-
    (w-1)^2 + c_{1,-} x_l + c_{a,-} x_{a^2} + c_{a,w,-} x_{a^2}(w-1) \nonumber \\
    & + &
    c_{a,a,-} x_{a^2}^2 + c_{a,m,-} x_l x_{a^2} + c_{2,-} x_l^2 \,\, .
    \nonumber
\end{eqnarray}
which depend on the light spectator quark mass $x_l = 2 B_0 \hat
m^\prime /(8 \pi^2 f_\pi^2)$ in the notation of
\cite{Bazavov:2011aa}, lattice spacing $x_{a^2} = [a/(4 \pi f_\pi
  r_1^2)]^2$, and $w = v\cdot v^\prime$.  The chiral ``logs'' term comes
from a staggered fermion version of the one-loop continuum result of
Chow and Wise \cite{Chow:1993hr} that includes taste-breaking
discretization effects \cite{Bailey:2012rr}.  We supplement the
next-to-next-leading order (NNLO) heavy-light meson staggered $\chi$PT
expression with terms analytic in $(w-1)$ to enable an interpolation
in $w$ at nonzero recoil.
%Thus, these fit functions contain the correct form at NLO in chiral
%perturbation theory, including staggered discretization effects.
The fit at NLO is already satisfactory ($p = 0.27$). The analytic NNLO
terms are then added with priors $0\pm 1$ so the final statistical
error contains the error of truncation.

The several sources of systematic error in the lattice determination
of $h_+$ and $h_-$ are listed in Table~\ref{tab:hphmsysbudget}
together with estimates of their contributions.  Data were adjusted to
our best tuned $\kappa$s, based on a calculation on one ensemble that
varied them.  The error listed reflects the uncertainty in the
adjustment.  We used $r_1 = 0.3117(22)$ fm.  The uncertainty in this
value is systematic.  The heavy-quark error in $h_+$ is estimated from
heavy quark effective theory with
$\mathcal{O}(\alpha_s(\Lambda/2m_Q)^2)$, and in $h_-$, with
$\mathcal{O}(\alpha_s\Lambda/2m_Q)$.  The current renormalization
factor $\rho^{V_4}$ is known to one loop. The error shown comes from
our estimate of the omitted higher order terms.  The corresponding
$\rho^{V_1}$ is not known.  We assume, very conservatively, that it
differs from unity by at most 20\%.  Because $h_-$ is about one tenth
the size of $h_+$, this choice does not impair the overall precision.

\begin{table}
    \caption{Systematic error budget (in percent). The total error is
      obtained by adding the individual errors in quadrature.  Not
      explicitly shown because they are negligible are finite-volume
      effects, isospin-splitting effects, and light-quark mass
      tuning.
    \label{tab:hphmsysbudget}
}
    \begin{center}
        \begin{tabular}{lcc}
            \hline
            source                                & $h_+$(\%) & $h_-$(\%)   \\
            \hline
             $\kappa$-tuning adjustment           & $\leq0.1$ & $1.4$ \\
             Lattice scale $r_1$                  & 0.2       & $\leq0.1$\\
             Heavy quark discretization           & 2.0       & 10. \\
             $\rho$ factor (current matching)     & 0.4       & 20.  \\
            \hline
            Total systematic error                & 2.1       &  22.    \\
            \hline
        \end{tabular}
    \end{center}
\end{table}

These errors are applied to the result of the chiral/continuum
extrapolation in quadrature with the statistical error.  The
thus-combined errors in $h_+$ and $h_-$ propagate to $f_+$ and $f_0$
according to the linear transformation Eq.~(\ref{eq:f_from_h}).  For a
more complete discussion, see \cite{B2Dpaper}.

\begin{figure}
\centerline{
  \includegraphics[width=0.5\textwidth]{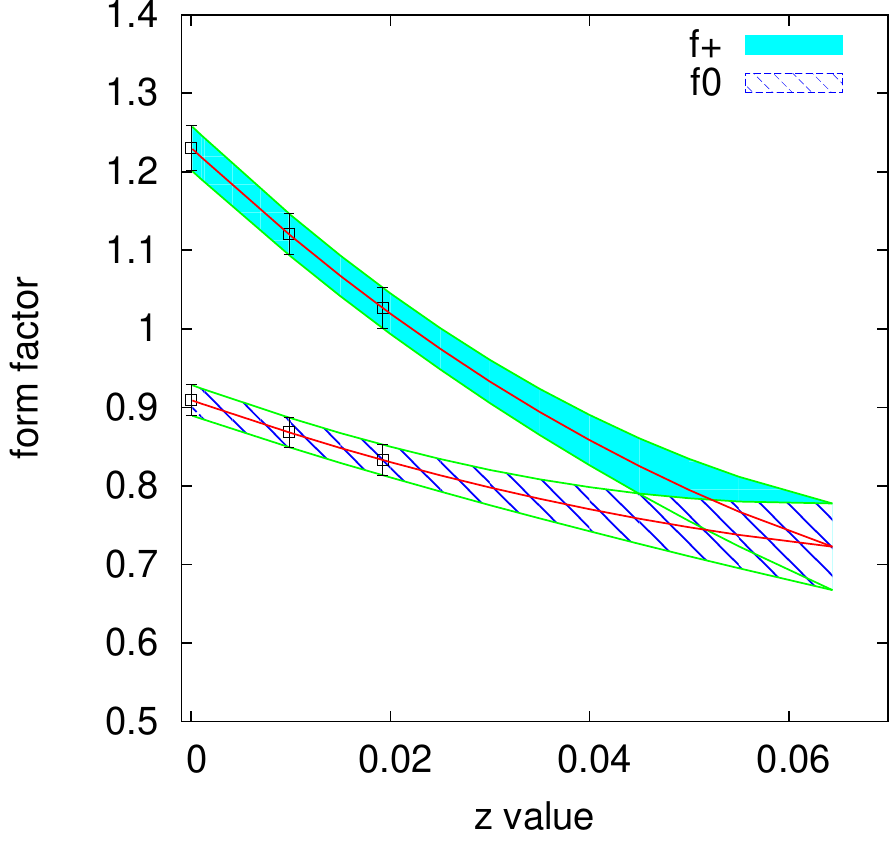}  \hfill
  \includegraphics[width=0.5\textwidth]{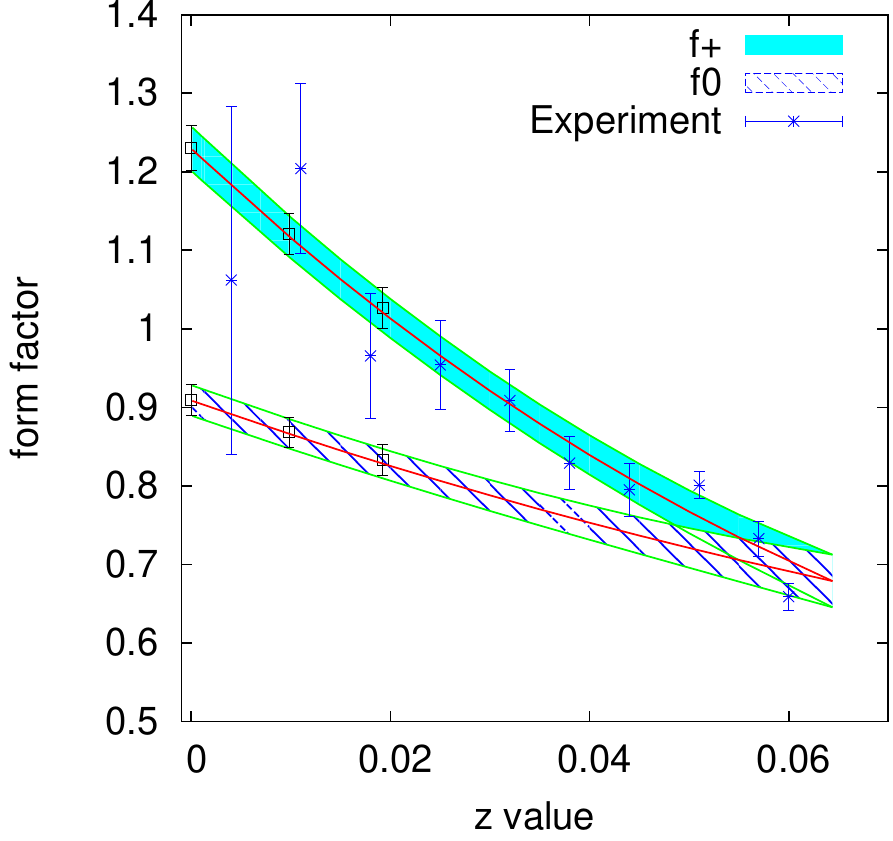}
}
  \caption{Left: form factors $f_+$ and $f_0$ parameterized by the $z$
    expansion to cubic order ($p = 0.12$).  Right: combined fit result
    with experimental results from the Babar collaboration
    \cite{Aubert:2009ac} ($p = 0.3$). The plotted experimental points
    have been divided by our best fit value of $\bar\eta_{EW}|V_{cb}|$
    and converted to $f_+$.  }
\label{fig:z-comparison}
\end{figure}

To compare the lattice and experimental form factors we need to
interpolate/extrapolate to a larger $w$ range.  We do this using the
$z$-expansion of Boyd, Grinstein and Lebed \cite{Boyd:1994tt}, which
provides a model-independent parameterization of the $q^2$ dependence
of $f_+$ and $f_0$.  This expansion builds in constraints from
analyticity and unitarity.  It is based on the conformal map
\begin{equation}
   z(w) = \frac{\sqrt{1+w}-\sqrt{2}}{\sqrt{1+w}+\sqrt{2}} \, ,
\end{equation}
which for \BDlnu\ maps the physical region $w \in [1,
  1.59]$ to $z \in [0, 0.0644]$.  It pushes poles and branch cuts far
away at $|z| \approx 1$.  Form factors are then parameterized as
\begin{equation}
  f_i(z) = \frac{1}{P_i(z) \phi_i(z)} \sum_{n=0}^N a_{i,n} z^n \, ,
\label{eq:zexp}
\end{equation}
where $N$ is the truncation order, $P_i(z)$ are the Blaschke factors
and $\phi_i$ are the ``outer functions''.  The latter are chosen to
simplify the unitarity bound: $\sum_n |a_{i,n}|^2 \le 1$.
The constraint on the sum of the coefficients, combined with the small
range of $z$, imply that we need only the first few coefficients in
the expansion.  We also impose the kinematic constraint $f_+ = f_0$ at
$q^2 = 0$ or $z \approx 0.0644$.

To implement the $z$ expansion, we start from the results for $f_+$
and $f_0$ at the continuum physical point, as determined from the
chiral/continuum fit.  We choose three $w$ values, $w=1.00$, 1.08, and
1.16, and fit the corresponding form factor data to determine the
coefficients $a_{i,j}$.  These, then, are used to parameterize the
form factors over the full kinematic range, as shown in the left panel
of Fig.~\ref{fig:z-comparison}.  We tested the truncation of the
$z$-expansion by adding higher terms with priors set to $0\pm1$ and
found the results were already stable at $N=2$.  We quote results for
$N=3$ so that the fit error incorporates the systematic error of
truncation.  Details of the fit result, including parameters and
correlations will be presented in the forthcoming paper \cite{B2Dpaper}.

When fitting to the experimental data it is necessary to take into
account electroweak effects still present in the experimental values,
but not included in the lattice calculation.  These include a Sirlin
factor $\eta_{\rm EW} = 1.00662$ for the $W\gamma$ and $WZ$ box
diagrams \cite{Atwood:1989em} and a further Coulomb correction for
final state interactions in $B^0$ decays.  BaBar reports that 37\% of
the decays were $B^0$s, which results in a QED correction factor in
the amplitude of $1 + 0.37 \alpha/(2\pi)$.  We have assigned an
uncertainty of $\pm 0.005$ to this correction to account for omitted
electromagnetic effects at intermedicate distances.  The net factor
is, then, $\bar \eta_{\rm EW} = 1.011(5)$.  (We use a bar to denote
the EW/QED correction of a sample of neutral and charged $B$s.)  We
prefer to use ${\cal G}(w)$ to denote the purely hadronic form factor,
so in our notation $\bar\eta_{EW}|V_{cb}|{\cal G}(w)$ corresponds to
the quantity often reported as $|V_{cb}|{\cal G}(w)$, and the ratio of
experimental to theoretical values must be divided by $\bar \eta_{\rm
  EW}$ to get $|V_{cb}|$.

The joint fit to our theoretical data and the experimental data from
the BaBar collaboration \cite{Aubert:2009ac} is shown in
Fig.~\ref{fig:z-comparison}.  The errors here include statistical and
systematic errors, combined in quadrature.  For the experimental
systematic error we assumed, for want of more accurate information,
that the quoted percentage value at small $w$ is appropriate over the
entire fit range \cite{Rotondo}.

\section{Results and discussion}

Our best fit value for $|V_{cb}|$ from the exclusive process \BDlnu\ at
nonzero recoil is
\begin{equation}
 |V_{cb}|=0.0385(19)_{\rm exp+lat}(2)_{\rm QED} \, .  % 11/29/13
\end{equation}
This value includes the full electroweak/QED correction.  The first
error combines statistical and systematic errors from both experiment
and theory.  The second error reflects the uncertainty in the Coulomb
correction. To get a sense of the relative importance of the various
systematic errors, we repeated the fit with only statistical errors
from both theory and experiment with the result 0.0388(11)(2).  With %  11/30/13
all errors, except the theoretical systematic errors, the result is
0.0383(17)(2).  With all errors, except the experimental systematic  %  11/30/13
errors, it is 0.0388(14)(2).  Thus we conclude that the experimental %  11/30/13
systematic error contributes more to the resulting error than the
theoretical one.

To quantify the improvement due to working at nonzero recoil, we also
use the standard method for extracting $|V_{cb}|$ based on
extrapolating the experimental data to zero recoil and comparing with
the theoretical form factor at this point. If we use the BaBar
collaboration result from the same BaBar data as in the nonzero recoil
analysis, namely, $\bar\eta_{EW}|V_{cb}| {\cal G}(1) = 0.0430(19)_{\rm
  stat}(14)_{\rm sys}$ \cite{Aubert:2009ac}, and our extrapolated
value ${\cal G}(1) = 1.081(25)$, we obtain           %  11/30/13
$|V_{cb}| = 0.0393(22)_{\rm exp+lat}(2)_{\rm QED}$.  %  11/30/13
So although the result is consistent with our nonzero recoil
determination, the error is 25\% larger.

A still better approach would be to fit the lattice results together
with a world average of the measured values of $\bar\eta_{EW}|V_{cb}|
{\cal G}(w)$ as a function of recoil parameter $w$ (suitably binned),
rather than with data from a single experiment, as we have done.  To
our knowledge, such a compilation has not been done
\cite{Amhis:2012bh}, but it would be welcome.

% \begin{figure}[b]
%     \centerline{\includegraphics[width=0.4\textwidth]{figs/compare.pdf}}
%     \caption{Comparison between exclusive determinations and inclusive
%       determination of $|V_{cb}|$.}
%     \label{fig:inex}
% \end{figure}

In our companion study of $ B\rightarrow D^* l\nu $ at zero recoil,
\cite{B2Dstarpaper} we obtain ${\cal F}(1) = 0.906(4)_{\rm
  stat}(12)_{\rm sys}$ for the hadronic form factor.  Here, in keeping
with our new convention for \BDlnu\, we reserve the notation ${\cal
  F}(1)$ for the purely hadronic form factor.  The values in the HFAG
world average reported in their notation as $|V_{cb}|{\cal F}(1)$ are,
in ours, $\bar\eta_{EW}|V_{cb}|{\cal F}(1)$ \cite{Amhis:2012bh}.
Because of the different proportion of neutral $B$ decays in this
average, we use $\bar\eta_{EW} = 1.015$. The result is
\begin{equation}
 |V_{cb}| = 0.0390(5)_{\rm exp}(5) _{\rm lat}(2) _{\rm QED} \, .
\end{equation}
%
% Our results are compared in Fig.~\ref{fig:inex} with the
% Gambino-Schwanda determination, based on inclusive decays of the $B$
% to any final state with charm.  
A recent analysis of inclusive decays quotes $|V_{cb}| =
0.0424(9)_{\rm exp+thy}$ \cite{Gambino:2013rza}. It disagrees at % 11/29/13
$1.8\sigma$ from ours for the exclusive $D$ final state and by
nearly $3 \sigma$ from our more precise result from the exclusive
$D^*$ final state.

The error in our determination of $|V_{cb}|$ from \BDlnu\ could be
improved by repeating the analysis with a world average of
experimental form factors, suitably binned in $w$, and by improving
our understanding of the experimental systematic error at larger $w$.
Improvements in lattice results will come from future studies with
still better actions, better statistics, smaller lattice spacing, and
ensembles with physical light quark masses that eliminate the need for
a chiral extrapolation.

\acknowledgments

Computations for this work were carried out with resources provided by
the USQCD Collaboration, the National Energy Research Scientific
Computing Center and the Argonne Leadership Computing Facility, which
is funded by the Office of Science of the U.S. Department of Energy;
and with resources provided by the National Institute for
Computational Science and the Texas Advanced Computing Center, which
are funded through the National Science Foundation's Teragrid/XSEDE
Program. This work was supported in part by the U.S.\ Department of
Energy under grants DE-FG02-13ER42011 (A.X.K) and DE-FC02-12ER-41879
(C.D.) and the U.S.\ National Science Foundation under grants
PHY07-57333 and PHY10-67881 (C.D.) and PHY09-03571 (S.-W.Q.), and by
the URA Visitng Scholars Program (A.X.K.). J.L.\ was supported by the
STFC and by the Scottish Universities Physics Alliance. Fermilab is
operated by Fermi Research Alliance, LLC, under Contract
No.\ DE-AC02-07CH11359 with the United States Department of Energy.

\providecommand{\href}[2]{#2}

\end{document}